\documentclass[12pt]{article}
\usepackage{palatino}
\usepackage[margin=1in]{geometry}
\usepackage{graphicx}
\usepackage{amssymb}
\usepackage{amsmath}
\usepackage{epstopdf}
\usepackage{color}
\usepackage{subfiles}
\usepackage[ragged]{sidecap}
\sidecaptionvpos{figure}{t}
\usepackage{mdwlist}
\usepackage[compact]{titlesec}
\titlespacing{\part}{0pt}{0pt}{10pt}
\titlespacing{\section}{0pt}{5pt}{5pt}
\titlespacing{\subsection}{0pt}{5pt}{5pt}
\titlespacing{\subsubsection}{0pt}{*1}{*1}
\usepackage{enumitem}
\usepackage{tabulary}
\newcolumntype{K}[1]{>{\centering\arraybackslash}p{#1}}
\usepackage{titlesec}
\usepackage{breakurl}
\usepackage{wrapfig}
\usepackage{blindtext}
\titleformat{\part}{\Large\bfseries}{\thepart}{1em}{}
\usepackage{hyperref}
\usepackage{caption}
\usepackage{multicol}

\newcommand{\Neff}{\ensuremath{N_\mathrm{eff}}}

\definecolor{orange}{rgb}{1,0.3,0}

\newcommand{\commentout}[1]{{}}

\def\lsim{\raise-.75ex\hbox{$\buildrel<\over\sim$}}

\DeclareUrlCommand\email{\urlstyle{rm}}

\newcommand{\dsr}{{DSR}}

\begin{document}

\def\bibname{References}

\bibliographystyle{utphys}  

\raggedbottom

\pagenumbering{roman}

\parindent=0pt
\parskip=8pt
\setlength{\evensidemargin}{0pt}
\setlength{\oddsidemargin}{0pt}
\setlength{\marginparsep}{0.0in}
\setlength{\marginparwidth}{0.0in}
\marginparpush=0pt


\definecolor{shadecolor}{rgb}{0.8,0.90,0.95}

\pagenumbering{roman} 

  \begin{center}
   {\Large\bf
      CMB-S4 Decadal Survey APC White Paper\\
   }
  \end{center}
 
\begin{center}
Contributors and Endorsers
\end{center}
\tolerance=4000

\newcounter{affilcount}
\newcommand{\affil}[1]{\refstepcounter{affilcount}\label{#1}}
\affil{UCIrvine}
\affil{JohnsHopkinsUniversity}
\affil{UniversityofIllinoisatUrbana-Champaign}
\affil{SLAC}
\affil{StanfordUniversity}
\affil{OxfordUniversity}
\affil{UCBerkeley}
\affil{LawrenceBerkeleyNationalLaboratory}
\affil{RiceUniversity}
\affil{Fermilab}
\affil{UCSanDiego}
\affil{SISSA}
\affil{ArgonneNationalLaboratory}
\affil{HarvardUniversity}
\affil{UniversityofNewMexico}
\affil{UniversityofChicago}
\affil{AstroParticleandCosmologyLaboratory}
\affil{Caltech}
\affil{CornellUniversity}
\affil{UniversityofPennsylvania}
\affil{CenterforAstrophysicsHarvardandSmithsonian}
\affil{YaleUniversity}
\affil{UniversityofCincinnati}
\affil{CaliforniaInstituteofTechnology}
\affil{LMUMunich}
\affil{INAF}
\affil{ItalianALMARegionalCentre}
\affil{CITA}
\affil{InstitutdAstrophysiquedeParis}
\affil{UniversityofManchester}
\affil{ArizonaStateUniversity}
\affil{CenterforComputationalAstrophysicsFlatironInstitute}
\affil{RutgersUniversity}
\affil{CardiffUniversity}
\affil{DartmouthCollege}
\affil{UniversityofSussex}
\affil{InstituteofAstronomyandDAMTPUniversityofCambridge}
\affil{McGillUniversity}
\affil{NIST}
\affil{StockholmUniversity}
\affil{NASAGoddardSpaceFlightCenter}
\affil{PrincetonUniversity}
\affil{SimonFraserUniversity}
\affil{UniversityofSouthernCalifornia}
\affil{HaverfordCollege}
\affil{UniversityofBritishColumbia}
\affil{UniversityofColoradoBoulder}
\affil{UniversityofMinnesota}
\affil{UniversityofMichigan}
\affil{KEK}
\affil{PennsylvaniaStateUniversity}
\affil{InstituteforAdvancedStudy}
\affil{UniversityofToronto}
\affil{FloridaStateUniversity}
\affil{ColumbiaUniversity}
\affil{NationalResearchCouncilCanada}
\affil{UniversityofVictoria}
\affil{KavliIPMU}
\affil{FitBit}
\affil{UCDavis}
\affil{UniversityofPittsburgh}
\affil{AaltoUniversity}
\affil{JPL}
\affil{StonyBrookUniversity}
\affil{UniversityofTokyo}
\affil{UniversityofGroningen}
\affil{CEASaclay}
\affil{SouthernMethodistUniversity}
\affil{InstitutLagrangedeParis}
\affil{EuropeanSouthernObservatory}
\affil{PerimeterInstitute}
\affil{DunlapInstitute}
\affil{NationalTaiwanUniversity}
\affil{UniversityofMilano-Bicocca}
\affil{BrookhavenNationalLaboratory}
\affil{UCLA}
\affil{UniversityofMelbourne}
\affil{CaseWesternReserveUniversity}
\affil{UniversityofCambridge}
\affil{DescartesLab}
\affil{KyotoUniversity}
\affil{UniversityofWisconsinMadison}
\affil{UniversitadegliStudidiMilan}
\affil{LAL}
\affil{BrownUniversity}
\affil{MassachusettsInstituteofTechnology}
\affil{SyracuseUniversity}
Kevork Abazajian,\textsuperscript{\ref{UCIrvine}}
Graeme Addison,\textsuperscript{\ref{JohnsHopkinsUniversity}}
Peter Adshead,\textsuperscript{\ref{UniversityofIllinoisatUrbana-Champaign}}
Zeeshan Ahmed,\textsuperscript{\ref{SLAC}}
Steven W.~Allen,\textsuperscript{\ref{StanfordUniversity}}
David Alonso,\textsuperscript{\ref{OxfordUniversity}}
Marcelo Alvarez,\textsuperscript{\ref{UCBerkeley},\ref{LawrenceBerkeleyNationalLaboratory}}
Mustafa A.~Amin,\textsuperscript{\ref{RiceUniversity}}
Adam Anderson,\textsuperscript{\ref{Fermilab}}
Kam~S.\ Arnold,\textsuperscript{\ref{UCSanDiego}}
Carlo Baccigalupi,\textsuperscript{\ref{SISSA}}
Kathy Bailey,\textsuperscript{\ref{ArgonneNationalLaboratory}}
Denis Barkats,\textsuperscript{\ref{HarvardUniversity}}
Darcy Barron,\textsuperscript{\ref{UniversityofNewMexico}}
Peter S.~Barry,\textsuperscript{\ref{UniversityofChicago}}
James~G.\ Bartlett,\textsuperscript{\ref{AstroParticleandCosmologyLaboratory}}
Ritoban Basu Thakur,\textsuperscript{\ref{Caltech}}
Nicholas Battaglia,\textsuperscript{\ref{CornellUniversity}}
Eric Baxter,\textsuperscript{\ref{UniversityofPennsylvania}}
Rachel Bean,\textsuperscript{\ref{CornellUniversity}}
Chris Bebek,\textsuperscript{\ref{LawrenceBerkeleyNationalLaboratory}}
Amy~N.\ Bender,\textsuperscript{\ref{ArgonneNationalLaboratory}}
Bradford A.~Benson,\textsuperscript{\ref{Fermilab}}
Edo Berger,\textsuperscript{\ref{CenterforAstrophysicsHarvardandSmithsonian}}
Sanah Bhimani,\textsuperscript{\ref{YaleUniversity}}
Colin A.~Bischoff,\textsuperscript{\ref{UniversityofCincinnati}}
Lindsey Bleem,\textsuperscript{\ref{ArgonneNationalLaboratory}}
James~J.\ Bock,\textsuperscript{\ref{CaliforniaInstituteofTechnology}}
Sebastian Bocquet,\textsuperscript{\ref{LMUMunich}}
Kimberly Boddy,\textsuperscript{\ref{JohnsHopkinsUniversity}}
Matteo Bonato,\textsuperscript{\ref{INAF},\ref{ItalianALMARegionalCentre}}
J.~Richard Bond,\textsuperscript{\ref{CITA}}
Julian Borrill,\textsuperscript{\ref{LawrenceBerkeleyNationalLaboratory},\ref{UCBerkeley}}
François R.~Bouchet,\textsuperscript{\ref{InstitutdAstrophysiquedeParis}}
Michael L.~Brown,\textsuperscript{\ref{UniversityofManchester}}
Sean Bryan,\textsuperscript{\ref{ArizonaStateUniversity}}
Blakesley Burkhart,\textsuperscript{\ref{CenterforComputationalAstrophysicsFlatironInstitute},\ref{RutgersUniversity}}
Victor Buza,\textsuperscript{\ref{HarvardUniversity}}
Karen Byrum,\textsuperscript{\ref{ArgonneNationalLaboratory}}
Erminia Calabrese,\textsuperscript{\ref{CardiffUniversity}}
Victoria  Calafut,\textsuperscript{\ref{CornellUniversity}}
Robert Caldwell,\textsuperscript{\ref{DartmouthCollege}}
John E.~Carlstrom,\textsuperscript{\ref{UniversityofChicago}}
Julien Carron,\textsuperscript{\ref{UniversityofSussex}}
Thomas Cecil,\textsuperscript{\ref{ArgonneNationalLaboratory}}
Anthony Challinor,\textsuperscript{\ref{InstituteofAstronomyandDAMTPUniversityofCambridge}}
Clarence L.~Chang,\textsuperscript{\ref{ArgonneNationalLaboratory}}
Yuji Chinone,\textsuperscript{\ref{UCBerkeley}}
Hsiao-Mei Sherry Cho,\textsuperscript{\ref{SLAC}}
Asantha Cooray,\textsuperscript{\ref{UCIrvine}}
Thomas M.~Crawford,\textsuperscript{\ref{UniversityofChicago}}
Abigail Crites,\textsuperscript{\ref{Caltech}}
Ari  Cukierman,\textsuperscript{\ref{SLAC}}
Francis-Yan Cyr-Racine,\textsuperscript{\ref{HarvardUniversity}}
Tijmen de Haan,\textsuperscript{\ref{LawrenceBerkeleyNationalLaboratory}}
Gianfranco de Zotti,\textsuperscript{\ref{INAF}}
Jacques Delabrouille,\textsuperscript{\ref{AstroParticleandCosmologyLaboratory}}
Marcel Demarteau,\textsuperscript{\ref{ArgonneNationalLaboratory}}
Mark Devlin,\textsuperscript{\ref{UniversityofPennsylvania}}
Eleonora Di Valentino,\textsuperscript{\ref{UniversityofManchester}}
Matt Dobbs,\textsuperscript{\ref{McGillUniversity}}
Shannon Duff,\textsuperscript{\ref{NIST}}
Adriaan Duivenvoorden,\textsuperscript{\ref{StockholmUniversity}}
Cora Dvorkin,\textsuperscript{\ref{HarvardUniversity}}
William Edwards,\textsuperscript{\ref{LawrenceBerkeleyNationalLaboratory}}
Joseph Eimer,\textsuperscript{\ref{JohnsHopkinsUniversity}}
Josquin Errard,\textsuperscript{\ref{AstroParticleandCosmologyLaboratory}}
Thomas Essinger-Hileman,\textsuperscript{\ref{NASAGoddardSpaceFlightCenter}}
Giulio Fabbian,\textsuperscript{\ref{UniversityofSussex}}
Chang Feng,\textsuperscript{\ref{UniversityofIllinoisatUrbana-Champaign}}
Simone Ferraro,\textsuperscript{\ref{LawrenceBerkeleyNationalLaboratory}}
Jeffrey P.~Filippini,\textsuperscript{\ref{UniversityofIllinoisatUrbana-Champaign}}
Raphael Flauger,\textsuperscript{\ref{UCSanDiego}}
Brenna Flaugher,\textsuperscript{\ref{Fermilab}}
Aurelien A.~Fraisse,\textsuperscript{\ref{PrincetonUniversity}}
Andrei Frolov,\textsuperscript{\ref{SimonFraserUniversity}}
Nicholas Galitzki,\textsuperscript{\ref{UCSanDiego}}
Silvia Galli,\textsuperscript{\ref{InstitutdAstrophysiquedeParis}}
Ken Ganga,\textsuperscript{\ref{AstroParticleandCosmologyLaboratory}}
Martina Gerbino,\textsuperscript{\ref{ArgonneNationalLaboratory}}
Murdock Gilchriese,\textsuperscript{\ref{LawrenceBerkeleyNationalLaboratory}}
Vera Gluscevic,\textsuperscript{\ref{UniversityofSouthernCalifornia}}
Daniel Green,\textsuperscript{\ref{UCSanDiego}}
Daniel Grin,\textsuperscript{\ref{HaverfordCollege}}
Evan Grohs,\textsuperscript{\ref{UCBerkeley}}
Riccardo Gualtieri,\textsuperscript{\ref{UniversityofIllinoisatUrbana-Champaign}}
Victor Guarino,\textsuperscript{\ref{ArgonneNationalLaboratory}}
Jon E.~Gudmundsson,\textsuperscript{\ref{StockholmUniversity}}
Salman Habib,\textsuperscript{\ref{ArgonneNationalLaboratory}}
Gunther Haller,\textsuperscript{\ref{SLAC}}
Mark Halpern,\textsuperscript{\ref{UniversityofBritishColumbia}}
Nils W.~Halverson,\textsuperscript{\ref{UniversityofColoradoBoulder}}
Shaul Hanany,\textsuperscript{\ref{UniversityofMinnesota}}
Kathleen Harrington,\textsuperscript{\ref{UniversityofMichigan}}
Masaya Hasegawa,\textsuperscript{\ref{KEK}}
Matthew Hasselfield,\textsuperscript{\ref{PennsylvaniaStateUniversity}}
Masashi Hazumi,\textsuperscript{\ref{KEK}}
Katrin Heitmann,\textsuperscript{\ref{ArgonneNationalLaboratory}}
Shawn Henderson,\textsuperscript{\ref{SLAC}}
Jason W.~Henning,\textsuperscript{\ref{UniversityofChicago}}
J. Colin Hill,\textsuperscript{\ref{InstituteforAdvancedStudy}}
Ren\'{e}e Hlo\v{z}ek,\textsuperscript{\ref{UniversityofToronto}}
Gil Holder,\textsuperscript{\ref{UniversityofIllinoisatUrbana-Champaign}}
William Holzapfel,\textsuperscript{\ref{UCBerkeley}}
Johannes Hubmayr,\textsuperscript{\ref{NIST}}
Kevin M.~Huffenberger,\textsuperscript{\ref{FloridaStateUniversity}}
Michael Huffer,\textsuperscript{\ref{SLAC}}
Howard Hui,\textsuperscript{\ref{Caltech}}
Kent Irwin,\textsuperscript{\ref{StanfordUniversity}}
Bradley R.~Johnson,\textsuperscript{\ref{ColumbiaUniversity}}
Doug Johnstone,\textsuperscript{\ref{NationalResearchCouncilCanada},\ref{UniversityofVictoria}}
William C.~Jones,\textsuperscript{\ref{PrincetonUniversity}}
Kirit Karkare,\textsuperscript{\ref{UniversityofChicago}}
Nobuhiko Katayama,\textsuperscript{\ref{KavliIPMU}}
James Kerby,\textsuperscript{\ref{ArgonneNationalLaboratory}}
Sarah Kernovsky,\textsuperscript{\ref{FitBit}}
Reijo Keskitalo,\textsuperscript{\ref{LawrenceBerkeleyNationalLaboratory},\ref{UCBerkeley}}
Theodore Kisner,\textsuperscript{\ref{LawrenceBerkeleyNationalLaboratory},\ref{UCBerkeley}}
Lloyd Knox,\textsuperscript{\ref{UCDavis}}
Arthur Kosowsky,\textsuperscript{\ref{UniversityofPittsburgh}}
John Kovac,\textsuperscript{\ref{HarvardUniversity}}
Ely D.~Kovetz,\textsuperscript{\ref{JohnsHopkinsUniversity}}
Steve Kuhlmann,\textsuperscript{\ref{ArgonneNationalLaboratory}}
Chao-lin Kuo,\textsuperscript{\ref{StanfordUniversity}}
Nadine Kurita,\textsuperscript{\ref{SLAC}}
Akito Kusaka,\textsuperscript{\ref{LawrenceBerkeleyNationalLaboratory}}
Anne Lahteenmaki,\textsuperscript{\ref{AaltoUniversity}}
Charles R.~Lawrence,\textsuperscript{\ref{JPL}}
Adrian T.~Lee,\textsuperscript{\ref{UCBerkeley},\ref{LawrenceBerkeleyNationalLaboratory}}
Antony Lewis,\textsuperscript{\ref{UniversityofSussex}}
Dale Li,\textsuperscript{\ref{SLAC}}
Eric Linder,\textsuperscript{\ref{LawrenceBerkeleyNationalLaboratory}}
Marilena Loverde,\textsuperscript{\ref{StonyBrookUniversity}}
Amy Lowitz,\textsuperscript{\ref{UniversityofChicago}}
Mathew S. Madhavacheril,\textsuperscript{\ref{PrincetonUniversity}}
Adam Mantz,\textsuperscript{\ref{StanfordUniversity}}
Frederick Matsuda,\textsuperscript{\ref{UniversityofTokyo}}
Philip Mauskopf,\textsuperscript{\ref{ArizonaStateUniversity}}
Jeff McMahon,\textsuperscript{\ref{UniversityofMichigan}}
P.~Daniel Meerburg,\textsuperscript{\ref{UniversityofGroningen}}
Jean-Baptiste Melin,\textsuperscript{\ref{CEASaclay}}
Joel Meyers,\textsuperscript{\ref{SouthernMethodistUniversity}}
Marius Millea,\textsuperscript{\ref{InstitutLagrangedeParis}}
Joseph Mohr,\textsuperscript{\ref{LMUMunich}}
Lorenzo Moncelsi,\textsuperscript{\ref{Caltech}}
Tony Mroczkowski,\textsuperscript{\ref{EuropeanSouthernObservatory}}
Suvodip Mukherjee,\textsuperscript{\ref{InstitutdAstrophysiquedeParis}}
Moritz M\"{u}nchmeyer,\textsuperscript{\ref{PerimeterInstitute}}
Daisuke Nagai,\textsuperscript{\ref{YaleUniversity}}
Johanna Nagy,\textsuperscript{\ref{DunlapInstitute},\ref{UniversityofToronto}}
Toshiya  Namikawa,\textsuperscript{\ref{NationalTaiwanUniversity}}
Federico Nati,\textsuperscript{\ref{UniversityofMilano-Bicocca}}
Tyler Natoli,\textsuperscript{\ref{DunlapInstitute}}
Mattia Negrello,\textsuperscript{\ref{CardiffUniversity}}
Laura Newburgh,\textsuperscript{\ref{YaleUniversity}}
Michael D.~Niemack,\textsuperscript{\ref{CornellUniversity}}
Haruki Nishino,\textsuperscript{\ref{KEK}}
Martin Nordby,\textsuperscript{\ref{SLAC}}
Valentine Novosad,\textsuperscript{\ref{ArgonneNationalLaboratory}}
Paul O'Connor,\textsuperscript{\ref{BrookhavenNationalLaboratory}}
Georges Obied,\textsuperscript{\ref{HarvardUniversity}}
Stephen Padin,\textsuperscript{\ref{UniversityofChicago}}
Shivam Pandey,\textsuperscript{\ref{UniversityofPennsylvania}}
Bruce Partridge,\textsuperscript{\ref{HaverfordCollege}}
Elena Pierpaoli,\textsuperscript{\ref{UniversityofSouthernCalifornia}}
Levon Pogosian,\textsuperscript{\ref{SimonFraserUniversity}}
Clement Pryke,\textsuperscript{\ref{UniversityofMinnesota}}
Giuseppe Puglisi,\textsuperscript{\ref{StanfordUniversity}}
Benjamin Racine,\textsuperscript{\ref{HarvardUniversity}}
Srinivasan Raghunathan,\textsuperscript{\ref{UCLA}}
Alexandra Rahlin,\textsuperscript{\ref{Fermilab}}
Srini Rajagopalan,\textsuperscript{\ref{BrookhavenNationalLaboratory}}
Marco Raveri,\textsuperscript{\ref{UniversityofChicago}}
Mark Reichanadter,\textsuperscript{\ref{SLAC}}
Christian L.~Reichardt,\textsuperscript{\ref{UniversityofMelbourne}}
Mathieu Remazeilles,\textsuperscript{\ref{UniversityofManchester}}
Graca Rocha,\textsuperscript{\ref{JPL}}
Natalie A.~Roe,\textsuperscript{\ref{LawrenceBerkeleyNationalLaboratory}}
Anirban Roy,\textsuperscript{\ref{SISSA}}
John Ruhl,\textsuperscript{\ref{CaseWesternReserveUniversity}}
Maria Salatino,\textsuperscript{\ref{AstroParticleandCosmologyLaboratory}}
Benjamin Saliwanchik,\textsuperscript{\ref{YaleUniversity}}
Emmanuel Schaan,\textsuperscript{\ref{LawrenceBerkeleyNationalLaboratory}}
Alessandro Schillaci,\textsuperscript{\ref{Caltech}}
Marcel M.~Schmittfull,\textsuperscript{\ref{InstituteforAdvancedStudy}}
Douglas Scott,\textsuperscript{\ref{UniversityofBritishColumbia}}
Neelima Sehgal,\textsuperscript{\ref{StonyBrookUniversity}}
Sarah Shandera,\textsuperscript{\ref{PennsylvaniaStateUniversity}}
Christopher Sheehy,\textsuperscript{\ref{BrookhavenNationalLaboratory}}
Blake~D.\ Sherwin,\textsuperscript{\ref{UniversityofCambridge}}
Erik Shirokoff,\textsuperscript{\ref{UniversityofChicago}}
Sara M.~Simon,\textsuperscript{\ref{UniversityofMichigan}}
An\v{z}e Slosar,\textsuperscript{\ref{BrookhavenNationalLaboratory}}
Rachel Somerville,\textsuperscript{\ref{CenterforComputationalAstrophysicsFlatironInstitute},\ref{RutgersUniversity}}
Suzanne T.~Staggs,\textsuperscript{\ref{PrincetonUniversity}}
Antony Stark,\textsuperscript{\ref{HarvardUniversity}}
Radek Stompor,\textsuperscript{\ref{AstroParticleandCosmologyLaboratory}}
Kyle T.~Story,\textsuperscript{\ref{DescartesLab}}
Chris Stoughton,\textsuperscript{\ref{Fermilab}}
Aritoki Suzuki,\textsuperscript{\ref{LawrenceBerkeleyNationalLaboratory}}
Osamu Tajima,\textsuperscript{\ref{KyotoUniversity}}
Grant P.~Teply,\textsuperscript{\ref{UCSanDiego}}
Keith Thompson,\textsuperscript{\ref{StanfordUniversity}}
Peter Timbie,\textsuperscript{\ref{UniversityofWisconsinMadison}}
Maurizio Tomasi,\textsuperscript{\ref{UniversitadegliStudidiMilan}}
Jesse I.~Treu,\textsuperscript{\ref{PrincetonUniversity}}
Matthieu Tristram,\textsuperscript{\ref{LAL}}
Gregory Tucker,\textsuperscript{\ref{BrownUniversity}}
Caterina Umiltà,\textsuperscript{\ref{UniversityofCincinnati}}
Alexander van Engelen,\textsuperscript{\ref{CITA}}
Joaquin D.~Vieira,\textsuperscript{\ref{UniversityofIllinoisatUrbana-Champaign}}
Abigail G.~Vieregg,\textsuperscript{\ref{UniversityofChicago}}
Mark Vogelsberger,\textsuperscript{\ref{MassachusettsInstituteofTechnology}}
Gensheng Wang,\textsuperscript{\ref{ArgonneNationalLaboratory}}
Scott Watson,\textsuperscript{\ref{SyracuseUniversity}}
Martin White,\textsuperscript{\ref{LawrenceBerkeleyNationalLaboratory},\ref{UCBerkeley}}
Nathan Whitehorn,\textsuperscript{\ref{UCLA}}
Edward J.\ Wollack,\textsuperscript{\ref{NASAGoddardSpaceFlightCenter}}
W.~L.~Kimmy Wu,\textsuperscript{\ref{UniversityofChicago}}
Zhilei Xu,\textsuperscript{\ref{UniversityofPennsylvania}}
Siavash Yasini,\textsuperscript{\ref{UniversityofSouthernCalifornia}}
James Yeck,\textsuperscript{\ref{UniversityofWisconsinMadison}}
Ki Won Yoon,\textsuperscript{\ref{StanfordUniversity}}
Edward Young,\textsuperscript{\ref{SLAC}}
Andrea Zonca\textsuperscript{\ref{UCSanDiego}}

\begin{multicols}{2}
\scriptsize
\setlength{\parskip}{2pt}

\noindent\textsuperscript{\ref{UCIrvine}}UC Irvine

\noindent\textsuperscript{\ref{JohnsHopkinsUniversity}}Johns Hopkins University

\noindent\textsuperscript{\ref{UniversityofIllinoisatUrbana-Champaign}}University of Illinois at Urbana-Champaign

\noindent\textsuperscript{\ref{SLAC}}SLAC

\noindent\textsuperscript{\ref{StanfordUniversity}}Stanford University

\noindent\textsuperscript{\ref{OxfordUniversity}}Oxford University

\noindent\textsuperscript{\ref{UCBerkeley}}UC Berkeley

\noindent\textsuperscript{\ref{LawrenceBerkeleyNationalLaboratory}}Lawrence Berkeley National Laboratory

\noindent\textsuperscript{\ref{RiceUniversity}}Rice University

\noindent\textsuperscript{\ref{Fermilab}}Fermilab

\noindent\textsuperscript{\ref{UCSanDiego}}UC San Diego

\noindent\textsuperscript{\ref{SISSA}}SISSA

\noindent\textsuperscript{\ref{ArgonneNationalLaboratory}}Argonne National Laboratory

\noindent\textsuperscript{\ref{HarvardUniversity}}Harvard University

\noindent\textsuperscript{\ref{UniversityofNewMexico}}University of New Mexico

\noindent\textsuperscript{\ref{UniversityofChicago}}University of Chicago

\noindent\textsuperscript{\ref{AstroParticleandCosmologyLaboratory}}AstroParticle \& Cosmology Laboratory

\noindent\textsuperscript{\ref{Caltech}}Caltech

\noindent\textsuperscript{\ref{CornellUniversity}}Cornell University

\noindent\textsuperscript{\ref{UniversityofPennsylvania}}University of Pennsylvania

\noindent\textsuperscript{\ref{CenterforAstrophysicsHarvardandSmithsonian}}Center for Astrophysics, Harvard \& Smithsonian

\noindent\textsuperscript{\ref{YaleUniversity}}Yale University

\noindent\textsuperscript{\ref{UniversityofCincinnati}}University of Cincinnati

\noindent\textsuperscript{\ref{CaliforniaInstituteofTechnology}}California Institute of Technology

\noindent\textsuperscript{\ref{LMUMunich}}LMU Munich

\noindent\textsuperscript{\ref{INAF}}INAF

\noindent\textsuperscript{\ref{ItalianALMARegionalCentre}}Italian ALMA Regional Centre

\noindent\textsuperscript{\ref{CITA}}CITA

\noindent\textsuperscript{\ref{InstitutdAstrophysiquedeParis}}Institut d'Astrophysique de Paris

\noindent\textsuperscript{\ref{UniversityofManchester}}University of Manchester

\noindent\textsuperscript{\ref{ArizonaStateUniversity}}Arizona State University

\noindent\textsuperscript{\ref{CenterforComputationalAstrophysicsFlatironInstitute}}Center for Computational Astrophysics, Flatiron Institute

\noindent\textsuperscript{\ref{RutgersUniversity}}Rutgers University

\noindent\textsuperscript{\ref{CardiffUniversity}}Cardiff University

\noindent\textsuperscript{\ref{DartmouthCollege}}Dartmouth College

\noindent\textsuperscript{\ref{UniversityofSussex}}University of Sussex

\noindent\textsuperscript{\ref{InstituteofAstronomyandDAMTPUniversityofCambridge}}Institute of Astronomy and DAMTP, University of Cambridge

\noindent\textsuperscript{\ref{McGillUniversity}}McGill University

\noindent\textsuperscript{\ref{NIST}}NIST

\noindent\textsuperscript{\ref{StockholmUniversity}}Stockholm University

\noindent\textsuperscript{\ref{NASAGoddardSpaceFlightCenter}}NASA Goddard Space Flight Center

\noindent\textsuperscript{\ref{PrincetonUniversity}}Princeton University

\noindent\textsuperscript{\ref{SimonFraserUniversity}}Simon Fraser University

\noindent\textsuperscript{\ref{UniversityofSouthernCalifornia}}University of Southern California

\noindent\textsuperscript{\ref{HaverfordCollege}}Haverford College

\noindent\textsuperscript{\ref{UniversityofBritishColumbia}}University of British Columbia

\noindent\textsuperscript{\ref{UniversityofColoradoBoulder}}University of Colorado Boulder

\noindent\textsuperscript{\ref{UniversityofMinnesota}}University of Minnesota

\noindent\textsuperscript{\ref{UniversityofMichigan}}University of Michigan

\noindent\textsuperscript{\ref{KEK}}KEK

\noindent\textsuperscript{\ref{PennsylvaniaStateUniversity}}Pennsylvania State University

\noindent\textsuperscript{\ref{InstituteforAdvancedStudy}}Institute for Advanced Study

\noindent\textsuperscript{\ref{UniversityofToronto}}University of Toronto

\noindent\textsuperscript{\ref{FloridaStateUniversity}}Florida State University

\noindent\textsuperscript{\ref{ColumbiaUniversity}}Columbia University

\noindent\textsuperscript{\ref{NationalResearchCouncilCanada}}National Research Council Canada

\noindent\textsuperscript{\ref{UniversityofVictoria}}University of Victoria

\noindent\textsuperscript{\ref{KavliIPMU}}Kavli IPMU

\noindent\textsuperscript{\ref{FitBit}}FitBit

\noindent\textsuperscript{\ref{UCDavis}}UC Davis

\noindent\textsuperscript{\ref{UniversityofPittsburgh}}University of Pittsburgh

\noindent\textsuperscript{\ref{AaltoUniversity}}Aalto University

\noindent\textsuperscript{\ref{JPL}}JPL

\noindent\textsuperscript{\ref{StonyBrookUniversity}}Stony Brook University

\noindent\textsuperscript{\ref{UniversityofTokyo}}University of Tokyo

\noindent\textsuperscript{\ref{UniversityofGroningen}}University of Groningen

\noindent\textsuperscript{\ref{CEASaclay}}CEA Saclay

\noindent\textsuperscript{\ref{SouthernMethodistUniversity}}Southern Methodist University

\noindent\textsuperscript{\ref{InstitutLagrangedeParis}}Institut Lagrange de Paris

\noindent\textsuperscript{\ref{EuropeanSouthernObservatory}}European Southern Observatory

\noindent\textsuperscript{\ref{PerimeterInstitute}}Perimeter Institute

\noindent\textsuperscript{\ref{DunlapInstitute}}Dunlap Institute

\noindent\textsuperscript{\ref{NationalTaiwanUniversity}}National Taiwan University

\noindent\textsuperscript{\ref{UniversityofMilano-Bicocca}}University of Milano-Bicocca

\noindent\textsuperscript{\ref{BrookhavenNationalLaboratory}}Brookhaven National Laboratory

\noindent\textsuperscript{\ref{UCLA}}UCLA

\noindent\textsuperscript{\ref{UniversityofMelbourne}}University of Melbourne

\noindent\textsuperscript{\ref{CaseWesternReserveUniversity}}Case Western Reserve University

\noindent\textsuperscript{\ref{UniversityofCambridge}}University of Cambridge

\noindent\textsuperscript{\ref{DescartesLab}}Descartes Lab

\noindent\textsuperscript{\ref{KyotoUniversity}}Kyoto University

\noindent\textsuperscript{\ref{UniversityofWisconsinMadison}}University of Wisconsin--Madison

\noindent\textsuperscript{\ref{UniversitadegliStudidiMilan}}Università degli Studi di Milan

\noindent\textsuperscript{\ref{LAL}}LAL

\noindent\textsuperscript{\ref{BrownUniversity}}Brown University

\noindent\textsuperscript{\ref{MassachusettsInstituteofTechnology}}Massachusetts Institute of Technology

\noindent\textsuperscript{\ref{SyracuseUniversity}}Syracuse University

\end{multicols}

\def\as#1{[{\bf AS:} {\it #1}] }

\eject
\pagenumbering{arabic} 
\setcounter{page}{1}

\section*{CMB-S4 Overview and Context}

CMB-S4 is envisioned to be the ultimate ground-based cosmic microwave background experiment, 
crossing critical thresholds in our understanding of the origin and evolution of the Universe, from the highest energies at the dawn of time through the growth of structure to the present day. The CMB-S4 science case is spectacular: the search for primordial gravitational waves as predicted from inflation and the imprint of relic particles including neutrinos,
unique 
insights into dark energy and tests of gravity on large scales, elucidating the role of baryonic feedback on galaxy formation and evolution, opening up a window on the transient Universe at millimeter wavelengths, and even the exploration of the outer Solar System. 
The CMB-S4 sensitivity to primordial gravitational waves will probe physics at the highest energy
scales and cross a major theoretically motivated threshold in constraints on inflation. 
The CMB-S4 search for new light relic particles will shed light on the early Universe 10,000 times farther
back than current experiments can reach.
Finally, the CMB-S4 Legacy Survey covering 70\% of the sky with
unprecedented sensitivity and angular resolution from centimeter- to millimeter-wave observing bands
will have a profound and lasting impact on Astronomy and Astrophysics and provide a powerful complement to surveys at other wavelengths, such as LSST and WFIRST, and others yet to be imagined.
We emphasize that these critical thresholds cannot be reached without the level of community and 
agency investment and commitment required by CMB-S4. In particular, the CMB-S4 science goals
are out of the reach of any projected precursor experiment by a significant margin.

CMB-S4 is planned to be a joint NSF and DOE project, with the construction phase to be funded as an NSF MREFC project and a DOE HEP MIE project. An interim project office has been constituted and tasked with advancing the CMB-S4 project in the NSF MREFC Preliminary Design Phase and toward DOE Critical Decision CD-1.
Support for the office is being provided in part by DOE, and a funding proposal to the NSF MSRI-R1 program is pending.  DOE 
CD-0 is expected imminently and will be a major milestone for the project. 

CMB-S4 was recommended by the 2014 Particle Physics Project Prioritization Panel (P5) report {\it Building for Discovery: Strategic Plan for U.S. Particle Physics in the Global Context\/} and by the 2015 National Academies report {\it A Strategic Vision for NSF Investments in Antarctic and Southern Ocean Research}. The community further developed the science case in the 2016 {\it CMB-S4 Science Book\/} \cite{Abazajian:2016yjj} and surveyed the status of the technology in the 2017 {\it CMB-S4 Technology Book} \cite{TechBookarXiv170602464A}.  This work formed the foundation for the joint NSF-DOE Concept Definition Task Force (CDT), a subpanel of the Astronomy and Astrophysics Advisory Committee (AAAC), a FACA committee advising DOE, NASA, and NSF. The CDT report was enthusiastically accepted by the AAAC in October 2017.  

Building on the CDT report, the CMB-S4 Collaboration and the pre-Project Development Group composed of experienced project leaders drawn primarily from the national laboratories have produced the comprehensive document, {\it The CMB-S4 Science Case, Reference Design, and Project Plan} \cite{Abazajian:2019}, which we refer to here as the Decadal Survey Report (\dsr).  The material presented in this white paper has been extracted from the 
\dsr, and we encourage the reader to see the \dsr\ for more detail. It and numerous other reports, collaboration bylaws, workshop and working group wiki pages, email lists, and much more may be found at the website \url{http://CMB-S4.org}.

To achieve its transformational science goals, CMB-S4 requires an enormous increase in sensitivity over all current CMB experiments combined, and roughly an order-of-magnitude increase over any projected precursor experiment. A significant and unique feature of CMB-S4 from the outset has been the use of multiple sites, specifically combining the two best currently developed sites on Earth for millimeter-wave observing: the high Atacama Plateau in Chile and the geographical South Pole. The design of CMB-S4 exploits key features of the two sites, namely the ability to drill deep on a single small patch of the sky through an extraordinarily stable atmosphere from the South Pole, and the ability to survey up to 80\% of the sky from the exceptionally high and dry Atacama site. 

Current experimental efforts at these two sites are already being consolidated into two major precursor observatories to CMB-S4, the Simons Observatory (SO) and the South Pole Observatory (SPO), whose teams also make up the vast majority of the CMB-S4 collaboration. The timing of both of these observatories is well-aligned with CMB-S4, enabling them to act as valuable pathfinders for CMB-S4 by providing technical and scientific data that have informed and will continue to inform our design and operations. To this end, both have also provided Letters of Intent to share their technical and cost data with CMB-S4. Nonetheless, while both will make significant advances in key CMB science goals, they will still fall well short of the thresholds targeted by CMB-S4. For example, to match the sensitivity to primordial gravitational waves provided by the ultra-deep CMB-S4 survey, SPO would have to 
integrate for nearly 50 years; it would take SO a similar amount of time to match the sensitivity to light relics provided by the CMB-S4 deep and wide survey. 

From space, the LiteBIRD CMB satellite mission was recently selected by JAXA for launch in 2028 for a 3-year mission, concurrent with CMB-S4 operations. With its lower resolution but wider frequency coverage, LiteBIRD's science goals are distinct from but highly complementary to CMB-S4's, and we are already discussing the parameters of a possible Memorandum of Understanding to enable both experiments to enhance their reach using elements of the other's data.

In short, CMB-S4 will enable transformational science that cannot be achieved otherwise, the CMB-S4 concept has clear community and agency support, and the CMB-S4 collaboration and project are moving forward. CMB-S4 thus represents a unique and timely scientific opportunity.

\section*{Key Science Goals and Objectives}
\label{sec:keyScience} 

We have organized the rich and diverse set of CMB-S4 scientific goals into four themes: 
\begin{enumerate}
\setlength{\itemsep}{0mm} 
\setlength{\parsep}{0mm}
\item \textit{primordial gravitational waves and inflation};
\item \textit{the dark Universe};
\item \textit{mapping matter in the cosmos};
\item \textit{the time-variable millimeter-wave sky}. 
\end{enumerate}
The first two science themes relate to fundamental physics. The other two themes relate to the broader scientific opportunities made possible by a millimeter-wave survey of unprecedented depth and breadth. Here we briefly review the key high-level goals and refer the reader to the science case detailed in the \dsr\ and in the decadal survey science white papers referenced.
 
\paragraph{Primordial gravitational waves and inflation.}

We have a historic opportunity to open up a window to the primordial Universe \cite{swp-shandera}. If the predictions of some of the leading models for the origin of the hot big bang are borne out, CMB-S4 will detect the signature of primordial gravitational waves in the polarization pattern of the CMB. This detection would provide the first evidence for the quantization of gravity, reveal new physics at the energy scale of grand unified theories, and yield insight into the symmetries of nature.

The current leading scenario for the origin of structure in our Universe is cosmic inflation, a period of accelerated expansion prior to the hot big bang. During this epoch, quantum fluctuations were imprinted on all spatial scales in the Universe. These fluctuations seeded the density perturbations that developed into all the structure in the Universe today. While 
there are still viable alternative models for the early history of the Universe, 
the simplest models of inflation are exceptionally successful in describing the data. 

Tantalizingly, the observed scale dependence of the amplitude of density perturbations has quantitative implications for the amplitude of primordial gravitational waves, commonly parameterized by $r$, the ratio of fluctuation power in gravitational waves to that in density perturbations. All inflation models that naturally explain the observed deviation from scale invariance and that also have a characteristic scale equal
to or larger than the gravitational mass scale 
predict $r \gtrsim 0.001$. A well-motivated sub-class within this set of models
is detectable by CMB-S4 at 5$\sigma$.  
The observed departure from scale invariance is a potentially important clue that strongly motivates exploring down to $r = 10^{-3}$.  With an order of magnitude more detectors than precursor observations, and exquisite control of systematic errors, CMB-S4 will improve upon limits from pre-CMB-S4 observations by a factor of five to reach this target, allowing us to either detect primordial gravitational waves or 
rule out large classes of inflationary models and dramatically impact how we think about the theory.

\paragraph{The dark Universe.}
In the standard cosmological model, about 95\% of the energy density of the Universe is in dark matter and dark energy.  With CMB-S4 we can address numerous questions about these dark ingredients, such as: How is matter distributed on large scales? Does the dark matter have non-gravitational interactions with baryons? Are there additional unseen components beyond dark matter and dark energy? 

Light relic particles are one very well-motivated possibility for additional energy density, as additional light particles appear frequently and numerously in extensions to the standard model of particle physics \cite{swp-green}. For large regions of the unexplored parameter space in these models, the light particles are thermalized in the early Universe.
The Planck satellite has sensitivity to light particles that fell out of thermal equilibrium in the first $\simeq 50$ micro-seconds of the Universe. With CMB-S4 we can push back this frontier by over a factor of 10,000, to the first fractions of a nanosecond. 

The contribution of light relics to the energy density, often parameterized as the ``effective number of neutrino species,'' $\Neff$, leads to observable consequences in the CMB temperature and polarization anisotropy. Current data are only sensitive enough to detect additional relics that froze out after the quark-hadron transition, 
so CMB-S4's ability to probe times well before that transition is a major advance. Specifically CMB-S4 will constrain $\Delta \Neff < 0.06$ at 95\% C.L., achieving sensitivity to Weyl fermion and vector particles that froze out at temperatures a few hundred times higher than that of the QCD phase transition.

CMB-S4 will also enable a broader exploration of the dark Universe in combination with other probes, often significantly enhancing them by breaking their intrinsic degeneracies.  It will improve or detect various possibilities for the dark matter properties beyond the simplest cold dark matter models \cite{swp-gluscevic}. 
It will add to dark energy constraints through precision measurements of the primordial power spectrum,
through precision measurements of the lensing convergence power spectrum, through the CMB-lensing-derived mass calibration of galaxy clusters \cite{swp-mantz}, and through CMB lensing tomography \cite{swp-slosar1}.

\paragraph{Mapping matter in the cosmos.}
Observations indicate there is 
roughly five times more dark matter than baryonic matter 
and that most of the  baryonic matter is in the form of hot ionized gas rather
than cold gas or stars.
CMB-S4 will be able to map out 
normal and dark 
matter separately by measuring the fluctuations in the
total mass density (using gravitational lensing) and the ionized gas density (using Compton scattering).

Observations of gravitational lensing of the CMB are key to many CMB-S4 science goals.
CMB-S4 lensing data will lead to a precise two-dimensional map of the total matter distribution.
The statistical properties of this mass map will provide important constraints on
dark energy \cite{swp-slosar1}, 
modified gravity \cite{swp-slosar1}, and the neutrino masses \cite{swp-dvorkin}.
When combined with CMB-S4-derived or external catalogs of galaxies or galaxy clusters, 
this mass map can be used to ``weigh'' the galaxy or cluster samples. 
With galaxies, this can be done in a redshift-dependent or tomographic manner  
 out to redshifts as high as $z \sim 5$,  
 making possible new precision tests of cosmology and gravity.
With robust CMB-lensing-based cluster masses
at high redshift, the abundance of galaxy clusters can be used as an additional
probe of dark energy and neutrino masses. 

Most of the baryons in the late Universe are believed to be in a diffuse ionized plasma 
that is difficult to observe \cite{swp-cicone,swp-oppenheimer,swp-wang}. 
CMB-S4 will measure the effect of Compton scattering by this gas 
(the Sunyaev-Zeldovich or SZ effects), both the spectral distortion from hot electrons
(thermal SZ or  tSZ) and a general redshift or blueshift of the scattered
photons due to coherent bulk flows along the line of sight (kinematic SZ or kSZ). 
The nature
of the scattering makes the SZ effects independent of redshift. 
With a deep and wide survey covering a large amount of volume and
an ultra-deep survey imaging lower-mass clusters, CMB-S4 will be an 
effective probe of 
the crucial regime of $z \gtrsim 2$, 
when galaxy clusters were vigorously accreting new hot gas while at the same time
forming the bulk of their stars \cite{swp-overzier}. 
The CMB-S4 catalog 
will contain an order of magnitude more clusters at $z > 2$ than will be discovered with Stage 3 CMB
experiments \cite{swp-mantz,swp-dannerbauer}. 
CMB-S4 will also measure the diffuse tSZ signal everywhere on the sky and make a 
temperature-weighted map of 
ionized gas 
that can be used to measure the average
thermal pressure profiles 
around 
galaxies and groups of galaxies. 
CMB-S4 will make maps of the kSZ effect, which will be combined with data from 
other surveys to make maps of the projected electron density 
around samples of objects. 
Applications of these maps include 
measuring ionized gas as a function of radius, directly constraining
the impact of feedback from active galactic nuclei and supernovae on the 
intergalactic medium \cite{swp-battaglia} and
constraining theories of modified gravity with the bulk flow amplitude as a function of separation. 
Even without overlapping galaxy catalogs, the kSZ signal can be used to probe 
the epoch of reionization, in ways that are highly complementary to the 
measurements of the neutral gas that can be obtained with redshifted Ly-$\alpha$
and 21-cm studies \cite{swp-chang,swp-cooray,swp-hutter,swp-laplante}.

\paragraph{The time-variable millimeter-wave sky.}

There have been relatively few studies of the variable sky at millimeter wavelengths, with
only one systematic survey done to date (by a CMB experiment \cite{swp-holder}).  
A deep, wide, millimeter-wave survey with time-domain capability will provide key insights
into transient or burst events, moving sources such as Solar-System objects, and variable
sources such as AGN.

Targeted follow-up observations of gamma-ray bursts, core-collapse supernovae, tidal disruption events,
classical novae, X-ray binaries, and stellar flares have found that there are many transient events
with measured fluxes that would make them detectable by CMB-S4.  A systematic survey of the mm-wave sky with a
cadence of a day or two over a large fraction of the sky, combined with
an ultra-deep daily survey of a few percent of the sky, would be an excellent
complement to other transient surveys, filling a gap between
radio and optical searches \cite{swp-holder}. Gamma-ray burst afterglows are particular interesting targets as they peak at millimeter wavelengths and there is a possibility
of capturing mm-wave afterglows that have no corresponding gamma-ray trigger, either
from the geometry of relativistic beaming and/or from sources at 
very high redshift \cite{swp-holder}. Both are predicted theoretically but have never been detected.

Thermal emission from planets, 
dwarf planets, and a selection of asteroids has been measured at these wavelengths; 
since these
sources move across the sky they can be differentiated from the 
stationary extrasolar sky. 
CMB-S4 will provide a long well-sampled time baseline and a wavelength range that is 
well-suited for the detection of possible large objects in the outer Solar System.
These measurements will be highly complementary
to those using optical reflected light or thermal emission at infrared wavelengths.

CMB-S4 will play an active role in multi-messenger astronomy, providing a long
baseline with high-cadence sampling in both intensity and linear polarization over
a wide sky area.
For example, the IceCube event IC170922A is believed to be associated with a flaring gamma-ray state of the blazar TXS 0506+056. In December 2014, however, the same source appears to have had a neutrino luminosity at least 10 times larger with no associated gamma emission---and no data existed at other wavelengths.
Having high-cadence wide-field non-gamma-ray data will be critical to understand sources like this one.
Any similar source is likely to be included in 
CMB-S4's near-daily, high-signal-to-noise monitoring 
of the blazar population.
The wide-area nature of the survey will also make it straightforward
to search for gravitational wave sources, particularly for
sources that happen to be poorly localized and are challenging for other instruments.

\section*{Technical Overview}
\label{sec:techOverview}


The CMB-S4 collaboration and project have
developed a Reference Design that meets the measurement requirements and therefore can deliver the CMB-S4 science goals. 
The main components of the Reference Design are described in detail in the \dsr\ and summarized here.
The major components of the Reference Design are as follows:

\begin{itemize}

\item An ultra-deep survey covering 3\% of the sky, more if a gravitational-wave signal is detected, to be conducted over seven years
using: fourteen 0.55-m refractor small-aperture telescopes (SATs) at 155\,GHz and below and four 0.44-m SATs at 220/270\,GHz, 
with dichroic, horn-coupled superconducting transition-edge-sensor (TES) detectors in each SAT, measuring two of the eight targeted frequency bands between 
30 and 270\,GHz; and one 6-m class ``delensing" large-aperture telescope (LAT), 
equipped with detectors distributed over seven bands from 20 to 278\,GHz.
Measurements at degree angular scales and larger made using refractor telescopes with roughly 0.5-m apertures have been demonstrated to deliver high-fidelity, low-contamination polarization measurements at these scales.  The combination of the SATs with the 6-m LAT therefore provides low-resolution $B$-mode measurements with excellent control of systematic contamination, as well as the high-resolution measurements required for delensing.  The ultra-deep survey SATs and 6-m LAT are to be located at the South Pole to allow targeted observations of the single small-area field, with provisions to relocate a fraction of the SATs in Chile if, for example, a high level of $r$ is detected or unforeseen systematic issues are encountered.

The total detector count for the 18 SATs is 
153,232, 
with the majority of the detectors allocated to the 85 to 155\,GHz bands.  
The total number of science-grade 150-mm detector wafers required for 18 SATs is 
204.
The delensing LAT will have a total TES detector count of
114,432,
with the majority of the detectors allocated to the 95 to 150\,GHz bands. 
The total number of science-grade 150-mm diameter detector wafers required for this single LAT is 76.  

\item A deep and wide survey covering approximately 70\% of the sky to be conducted over seven years using two 6-m 
LATs located in Chile, each equipped with 
121,760
TES detectors distributed over eight frequency bands spanning 30 to 278\,GHz. The total number of science-grade 150-mm diameter detector wafers required for these two LATs is 152.
\end{itemize}

In the context of their legacy value to the wider community, we refer to the deep/wide and ultra-deep high-resolution surveys together as the CMB-S4 Legacy Survey.
The total detector count for CMB-S4 is 
511,184, 
requiring 
432
science grade wafers. This is an enormous increase over the detector count of all Stage-3 experiments combined. Such a dramatic increase in scale is required to meet the CMB-S4 science goals. 

\section*{Technical Drivers}

The CMB-S4 reference design uses existing, well-demonstrated technology that has been developed and demonstrated by the CMB experimental groups over the last decade, scaled up to unprecedented levels. The design and implementation plan addresses the considerable technical challenges presented by the required scaling up of the instrumentation and by the scope and complexity of the data analysis and interpretation.  Features of the design and plan include: scaled-up superconducting detector arrays with well-understood and robust material properties and processing techniques; high-throughput mm-wave telescopes and optics with unprecedented precision and rejection of systematic contamination; full internal characterization of astronomical foreground emission; large cosmological simulations and improved theoretical modeling; and computational methods for extracting minute correlations in massive, multi-frequency data sets, which include noise and a host of known and unknown signals. 

A CMB-S4 Risk and Opportunity Management Plan describes the continuous risk and opportunity management process implemented by the project, consistent with DOE O413.3B, “Project Management for the Acquisition of Capital Assets,” and the NSF 17-066, “NSF Large Facilities Manual.”  The plan establishes the methods of assessing CMB-S4 project risk and opportunities for all subsystems as well as the system as a whole. The CMB-S4 risk register has 213 risks identified. There are three risks that are currently assessed at Critical and 26 risks at High. The project is working on mitigations to ensure that these risks are lowered to reasonable levels on a timescale consistent with our overall project timeline. 

For example, a current identified critical risk is meeting the scaled-up production and testing timeline of the transition-edge-sensor detector arrays.  This is a major focus of the R\&D program supported by the DOE.  The Interim Project Office formed a Detector and Readout (D\&R) Task Force in early 2019 to evaluate existing fabrication and testing capabilities and to provide recommendations on production plans. A formal review of the resulting detector fabrication plan will be completed in mid-2019.

\section*{Organization, Partnerships, and Current Status}

CMB-S4 is both a scientific collaboration and a nascent DOE/NSF project. While these are certainly tightly coupled, they do have different roles and responsibilities; the overall organization of CMB-S4 therefore decouples into the organization of the collaboration and the project.

The formal CMB-S4 collaboration was established in 2018 with the ratification of the bylaws and election of the various officers including the collaboration Governing Board.
As of summer 2019 the collaboration has 198 members, 71 of whom hold positions within the organizational structure. These members represent 11 countries on 4 continents, and 76 institutions comprising 16 national laboratories and 60 universities.

The CMB-S4 Collaboration and a pre-Project Development Group of experienced project leaders  drawn largely from the national labs, jointly contributed to the development of the project Work Breakdown Structure (WBS), Organization, Cost Book, Resource Loaded Schedule, and Risk Registry.  The top level WBS Structure and Cost is summarized in Table~\ref{tbl:wbs_cost}. The schedule has 1110 activities, 1928 relationships, 5 Level 1, 20 Level 2 and 299 Level 3 Milestones for the CMB-S4 project.

\begin{figure}[htb]
\begin{center}
\includegraphics[width=0.98\textwidth]{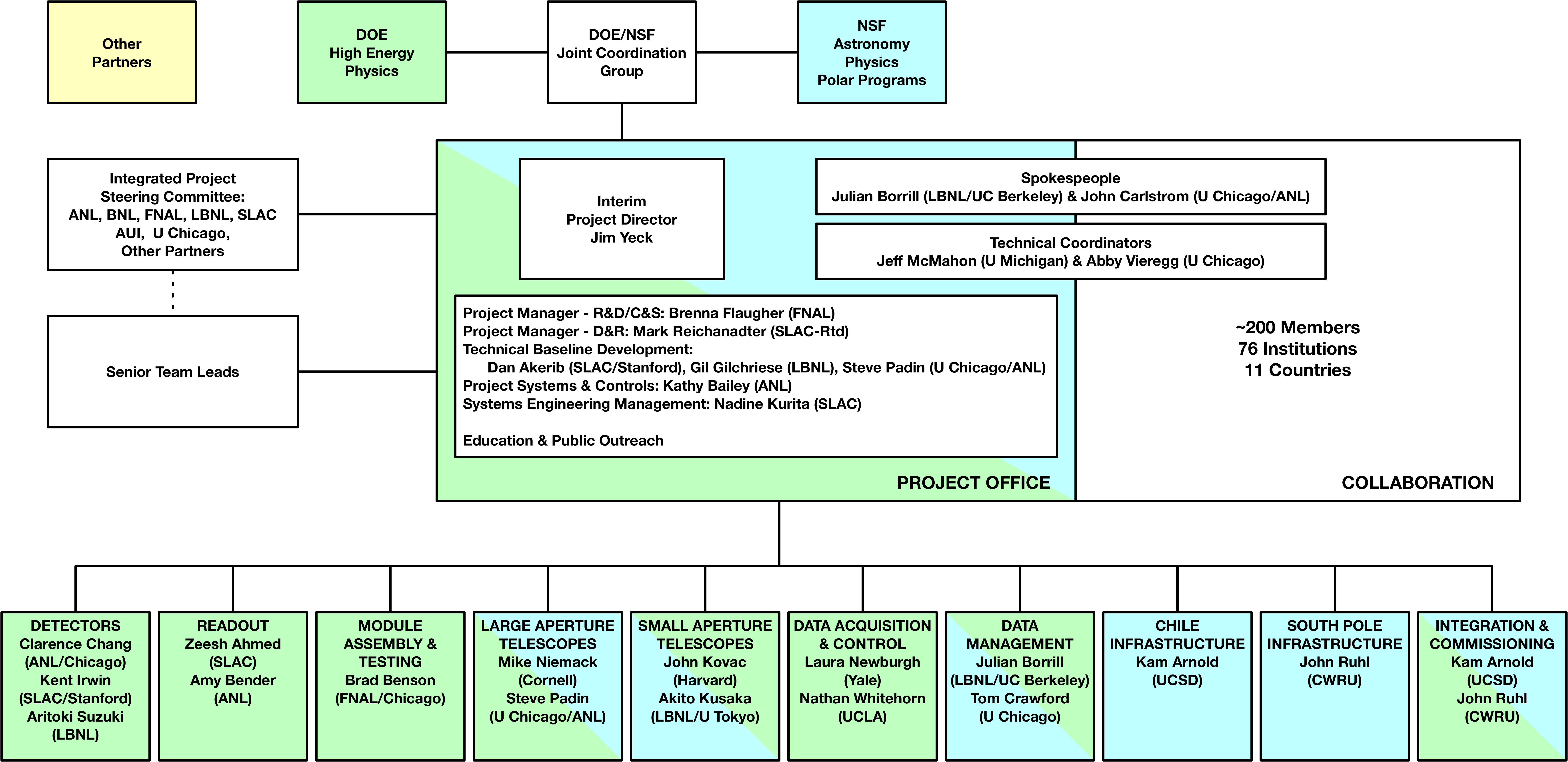}
\captionsetup{font=footnotesize}
\caption{Organizational Chart of the Interim Project Office.  The figure includes a notional distribution of project scope by funding agency (NSF = blue, DOE = green, Other = yellow).  We are actively pursuing partners who could make significant scope contributions in areas aligned with their expertise.}
\label{fig:prj_org}
\end{center}
\vskip -20pt
\end{figure}

The reference design and project baseline summarized here and detailed in the \dsr\  is the basis for subsequent design and project development work to be led by the Interim Project Office (see Fig.~\ref{fig:prj_org}) and the Collaboration.  A permanent Integrated Project Office will be established in 2020 to manage the construction phase which is anticipated to start in 2021.  

\begin{wraptable}{r}{3.5in}
\captionsetup{font=footnotesize}
\vspace{-.2in}
\caption{CMB-S4 WBS Structure and Cost}
\vspace{-0.25in}
{\footnotesize
\begin{center}
\begin{tabular}{|l|r|}
\hline
\multicolumn{1}{|c|}{WBS Level 2 Element} & \multicolumn{1}{|c|}{\$M} \\
\hline
\multicolumn{2}{|c|}{Total Estimated Cost (TEC)} \\
\hline
1.01 -- Project Management & 19.6\\
1.03 -- Detectors & 39.5 \\
1.04 -- Readout & 59.9 \\
1.05 -- Module Assembly \& Testing & 31.8 \\
1.06 -- Large Aperture Telescopes & 86.5 \\
1.07 -- Small Aperture Telescopes & 52.3   \\
1.08 -- Observation Control \& Data Acquisition & 13.9   \\
1.09 -- Data Management & 26.9   \\
1.10 -- Chile Infrastructure & 38.1   \\
1.11 -- South Pole Infrastructure & 37.0   \\
1.12 -- Integration \& Commissioning & 7.7   \\
\hline
Direct TEC & 413.2   \\
TEC Contingency (35\%) & 144.6   \\
Total TEC & 557.9   \\
\hline
\hline
\multicolumn{2}{|c|}{Other Project Cost (OPC)} \\
\hline
1.01 -- Project Management & 7.0   \\
1.02 -- Research \& Development & 24.2   \\
\hline
Direct OPC & 31.2   \\
OPC Contingency (35\%) - excludes R\&D & 2.5   \\
Total OPC & 33.7   \\
\hline
\hline
\multicolumn{2}{|c|}{Total Project Cost (TPC)} \\
\hline
TEC + OPC with contingency & 591.6   \\
\hline
\end{tabular}
\end{center}
\label{tbl:wbs_cost}
}
\vspace{-.4in}
\end{wraptable}
As shown in Fig.~\ref{fig:prj_org}, a key feature of the organization is the role of collaboration members in the project office, in particular as leaders of the Level 2 systems.  The Level 2 managers are supported by engineering and project-management leaders.  The NSF/DOE scope distribution will promote the engagement and participation of universities and national laboratories.  Graduate students, postdocs, professional technicians and engineers are expected to be involved in all aspects of the project.

The project office is responsible for forming partnerships with key stakeholder institutions, including DOE National Laboratories, universities, and potential collaborating projects such as the Simons Observatory, South Pole Observatory, and the CCAT-prime project.  Partnerships are also expected to include foreign institutions participating in the CMB-S4 Science Collaboration and contributing to the CMB-S4 Project.  

The CMB-S4 project is expected to include significant contributions from collaborating institutions supported by funding agencies other than NSF and DOE.  These ``in-kind'' contributions will be defined as deliverables to the project. Major contributions from partners will need to be negotiated and incorporated in the project design within the next two to three years to avoid adding schedule and cost risk.

\section*{Schedule} 

Table~\ref{tbl:schedule} shows the proposed timeline via the NSF Level 1 Milestones along with the corresponding DOE Critical Decision Milestones.
 The schedule development strategy is to define a schedule that is consistent with the funding potentially available during FY2019-FY2021 and is subsequently  technically driven.  The project is working towards an early completion milestone that contains limited schedule float. A year of schedule float following this early project complete milestone is included in the overall project complete milestone CD-4.  The Interim Project Office will continue to optimize the schedule and include explicit float for activities that are not on the critical path.  The best opportunity to improve on the schedule is to reduce the time required to deliver the full quantity of the Detectors and Readout (D\&R) components.  

Seven years of operations are needed to achieve the CMB-S4 science goals. 

\begin{table}[htp]
\captionsetup{font=footnotesize}
{\footnotesize
\caption{Timeline and Funding Agency Milestones}
\vspace{-0.2in}
\begin{center}
\begin{tabular}{|c|c|}
\hline
NSF Level 1 Milestone (DOE Critical Decision) & Schedule (FY) \\
\hline
Pre-Conceptual Design (CD-0, Mission Need) & Q3 2019 \\
Preliminary Baseline (CD-1/3a, Cost Range/Long-Lead Procurement) & Q3 2021 \\
Preliminary Design Review (CD-2, Performance Baseline) & Q2 2022 \\
Final Deign Review (CD-3, Start of Construction) & Q4 2023 \\
Completion of 1st Telescope (CD-4a, Initial Operations) & Q2 2026 \\
Project Completion(CD-4, Operations) & Q1 2029 \\
\hline
\end{tabular}
\end{center}
\label{tbl:schedule}
}
\vspace{-0.3in}
\end{table}

\section*{Cost Estimates}


The CMB-S4 project total estimated cost is currently \$591.6M (fully loaded and escalated to the year of expenditure) including a 35\% contingency budget.  The breakdown of the costs by major components of the construction phase is shown in Table~\ref{tbl:wbs_cost}.
The cost estimate is the full cost, i.e., it does not take credit for use of any legacy infrastructure or for contributions from collaborating institutions supported by private and international partners, e.g., large-aperture telescopes currently under construction in Chile as part of the Simons Observatory, or large- and small-aperture telescopes proposed by international collaborators.  In-kind contributions delivered by private and international partners are expected and would reduce the total cost to NSF and DOE.  It is estimated that the value of in-kind contributions could reduce the total cost of the CMB-S4 project by 20-25\%.

The total estimated cost is built on detailed cost estimates made for each task in the project schedule.  The estimates are documented with a Basis of Estimate (BOE) developed by the subsystems leads. The task resources and their quantities are assigned from a standardized list of resources. The list includes multiple resource classes in each of the categories: labor, materials/non-labor, or travel. A task estimate consists of the number of hours of each labor resource class, the base-year dollar cost of each materials/non-labor resource class, the number of trips for each travel resource class, and the basis for each estimate. 

The cost contingency estimate was constructed using input from experts with experience in previous CMB experiments and similar NSF MREFC projects and DOE MIE projects.  As the design, cost estimates, and schedules mature the contingency as a percentage of the base cost estimate is expected to decrease to 30\% or less.  The target range for the start of the CMB-S4 construction project is 25-30\%.
A notional distribution of project scope by funding agency is shown in Fig.~\ref{fig:prj_org}, where blue indicates NSF and green DOE.  
The level of the NSF and DOE costs are expected to be comparable, 
with the notional distribution having NSF and DOE contributing 42\% and 58\% of the funding, respectively.
 
The basic operations model for CMB-S4 will be observations with multiple telescopes and cameras distributed across two sites, with observing priorities and specifications optimized for the CMB-S4 science goals, and data from all instruments shared throughout the entire CMB-S4 collaboration.  The operations cost is based on a preliminary bottom-up estimate that includes management, site staff, utilities, instrument maintenance, data transmission, data products, pipeline upgrades, collaboration management, and key science analysis. The annual operations cost is \$32M in 2019 dollars, excluding 20 FTE/year of scientist effort supported by DOE research funds, with roughly 60\% allocated to operations and 40\% allocated to analysis of the key CMB-S4 science goals. 
The non-key science analysis of CMB-S4 data products will be carried out by laboratory and university scientists, with support for the latter expected to be provided by individual NSF and DOE awards.

Normal end-of-life decommissioning costs for South Pole and Chile infrastructure are anticipated.

\eject

\markboth{\bibname}{}
\markright{\bibname}
\bibliography{cmbs4}

\providecommand{\href}[2]{#2}\begingroup\raggedright\begin{thebibliography}{10}

\bibitem{Abazajian:2016yjj}
{\bfseries CMB-S4} Collaboration, K.~N. Abazajian {\em et~al.}, ``{CMB-S4
  Science Book, First Edition},''
\href{http://arxiv.org/abs/1610.02743}{{\ttfamily arXiv:1610.02743
  [astro-ph.CO]}}.

\bibitem{TechBookarXiv170602464A}
{\bfseries CMB-S4} Collaboration, M.~H. Abitbol {\em et~al.}, ``{CMB-S4
  Technology Book, First Edition},'' {\em arXiv e-prints} (Jun, 2017)
  arXiv:1706.02464, \href{http://arxiv.org/abs/1706.02464}{{\ttfamily
  arXiv:1706.02464 [astro-ph.IM]}}.

\bibitem{Abazajian:2019}
{\bfseries CMB-S4} Collaboration, K.~N. Abazajian {\em et~al.}, ``{CMB-S4
  Science Case, Reference Design, and Project Plan},''
  \href{http://arxiv.org/abs/1907.04473}{{\ttfamily arXiv:1907.04473
  [astro-ph.CO]}}.

\bibitem{swp-shandera}
S.~Shandera {\em et~al.}, ``Probing the origin of our universe through cosmic
  microwave background constraints on gravitational waves,''
  \href{http://arxiv.org/abs/Decadal Survey Science White Paper}{{\ttfamily
  Decadal Survey Science White Paper}}.

\bibitem{swp-green}
D.~Green {\em et~al.}, ``Messengers from the early universe: Cosmic neutrinos
  and other light relics,'' \href{http://arxiv.org/abs/Decadal Survey Science
  White Paper}{{\ttfamily Decadal Survey Science White Paper}}.

\bibitem{swp-gluscevic}
V.~Gluscevic {\em et~al.}, ``Cosmological probes of dark matter interactions:
  The next decade,'' \href{http://arxiv.org/abs/Decadal Survey Science White
  Paper}{{\ttfamily Decadal Survey Science White Paper}}.

\bibitem{swp-mantz}
A.~Mantz {\em et~al.}, ``The future landscape of high-redshift galaxy cluster
  science,'' \href{http://arxiv.org/abs/Decadal Survey Science White
  Paper}{{\ttfamily Decadal Survey Science White Paper}}.

\bibitem{swp-slosar1}
A.~Slosar {\em et~al.}, ``Dark energy and modified gravity,''
  \href{http://arxiv.org/abs/Decadal Survey Science White Paper}{{\ttfamily
  Decadal Survey Science White Paper}}.

\bibitem{swp-dvorkin}
C.~Dvorkin {\em et~al.}, ``Neutrino mass from cosmology: Probing physics beyond
  the standard model,'' \href{http://arxiv.org/abs/Decadal Survey Science White
  Paper}{{\ttfamily Decadal Survey Science White Paper}}.

\bibitem{swp-cicone}
C.~Cicone {\em et~al.}, ``The hidden circumgalactic medium,''
  \href{http://arxiv.org/abs/Decadal Survey Science White Paper}{{\ttfamily
  Decadal Survey Science White Paper}}.

\bibitem{swp-oppenheimer}
B.~Oppenheimer {\em et~al.}, ``Imprint of drivers of galaxy formation in the
  circumgalactic medium,'' \href{http://arxiv.org/abs/Decadal Survey Science
  White Paper}{{\ttfamily Decadal Survey Science White Paper}}.

\bibitem{swp-wang}
Q.~D. Wang {\em et~al.}, ``The panchromatic circumgalactic medium,''
  \href{http://arxiv.org/abs/Decadal Survey Science White Paper}{{\ttfamily
  Decadal Survey Science White Paper}}.

\bibitem{swp-overzier}
R.~Overzier {\em et~al.}, ``Tracing the formation history of galaxy clusters
  into the epoch of reionization,'' \href{http://arxiv.org/abs/Decadal Survey
  Science White Paper}{{\ttfamily Decadal Survey Science White Paper}}.

\bibitem{swp-dannerbauer}
H.~Dannerbauer {\em et~al.}, ``Mapping galaxy clusters in the distant
  universe,'' \href{http://arxiv.org/abs/Decadal Survey Science White
  Paper}{{\ttfamily Decadal Survey Science White Paper}}.

\bibitem{swp-battaglia}
N.~Battaglia {\em et~al.}, ``Probing feedback in galaxy formation with
  millimeter-wave observations,'' \href{http://arxiv.org/abs/Decadal Survey
  Science White Paper}{{\ttfamily Decadal Survey Science White Paper}}.

\bibitem{swp-chang}
T.-C. Chang {\em et~al.}, ``Tomography of the cosmic dawn and reionization eras
  with multiple tracers,'' \href{http://arxiv.org/abs/Decadal Survey Science
  White Paper}{{\ttfamily Decadal Survey Science White Paper}}.

\bibitem{swp-cooray}
A.~Cooray {\em et~al.}, ``Cosmic dawn and reionization: Astrophysics in the
  final frontier,'' \href{http://arxiv.org/abs/Decadal Survey Science White
  Paper}{{\ttfamily Decadal Survey Science White Paper}}.

\bibitem{swp-hutter}
A.~Hutter {\em et~al.}, ``A proposal to exploit galaxy-21cm synergies to shed
  light on the epoch of reionization,'' \href{http://arxiv.org/abs/Decadal
  Survey Science White Paper}{{\ttfamily Decadal Survey Science White Paper}}.

\bibitem{swp-laplante}
P.~L. Plante {\em et~al.}, ``Mapping cosmic dawn and reionization: Challenges
  and synergies,'' \href{http://arxiv.org/abs/Decadal Survey Science White
  Paper}{{\ttfamily Decadal Survey Science White Paper}}.

\bibitem{swp-holder}
G.~Holder {\em et~al.}, ``Tracking the time-variable millimeter-wave sky with
  cmb experiments,'' \href{http://arxiv.org/abs/Decadal Survey Science White
  Paper}{{\ttfamily Decadal Survey Science White Paper}}.

\end{thebibliography}\endgroup

\end{document}